\documentstyle[epsf]{ioplppt}                
\begin{document}
\letter{Chiral feedback for $p$-wave superconductors}

\author{J. Goryo\ftnote{1}{e-mail;goryo@yukawa.kyoto-u.ac.jp} and M. Sigrist}

\address{Yukawa Institute for Theoretical Physics, Kyoto University, 
Kyoto 606-8502, Japan}

\begin{abstract}
In a quasi-two-dimensional $ p $-wave superconductor there find six Cooper
pairing states which are degenerate within the weak-coupling
approach. We show that this degeneracy can be lifted by feedback effect
favoring the so-called chiral $ p $-wave state. This effect is based
on the anomalous coupling between charge and current in a system with
broken time reversal symmetry and parity. 
\end{abstract}


The discovery of odd-parity Cooper pairing in Sr$_2$RuO$_4$ led to the 
renewed interest in so-called p-wave (spin-triplet)
superconductivity.\cite{MAENO,ISHIDA} There is a number of other
systems, such as the heavy 
Fermion superconductors UPt$_3$ and UBe$_{13}$ or the organic
superconductor (TMTSF)$_2$PF$_6$, where very likely odd-parity pairing 
is realized.\cite{TOU,CHAIKIN}  The spin-1 degree of
freedom of the spin-triplet Cooper pairs provides a considerably wider
space of potential pairing states than in the even-parity (spin
singlet) case.\cite{REV} The symmetry and effective dimensionality  
of the electronic band structure plays an
important role in determining the possible p-wave pairing states. The
examples listed above cover many of the possibilities: the organic
superconductor is quasi-one-dimensional, the heavy Fermion compounds
are three-dimensional, while Sr$_2$RuO$_4$ represents the case of
quasi-two-dimensional system. 

The large number of possible p-wave states makes their identification 
for each material a difficult task. 
A simple weak-coupling BCS type of approach can give a first 
guess on the most stable state. Because the condensation energy is in
this case directly connected with the presence of the energy gap in the
quasiparticle spectrum, the state with the least nodes in the gap
would be most favorable. In one and three dimensions the most stable
state is unique up to spin rotation. Assuming parabolic
band structure for the corresponding dimension and using the d-vector
notation one finds
\begin{equation} \begin{array}{ll}
{\bf d} ({\bf k}) = \hat{{\bf x}} k_x & \qquad \mbox{for 1d }\\
{\bf d} ({\bf k}) = \hat{{\bf x}} k_x + \hat{{\bf y}} k_y +
\hat{{\bf z}} k_z & \qquad \mbox{for 3 d} 
\end{array} \end{equation}
where the gap matrix is defined as $ \hat{\Delta}_{{\bf k}} = i
\sigma_{2} {\boldmath \sigma} \cdot {\bf d}({\bf k}) $ and the
quasiparticle gap is $ \frac{1}{2} tr [ \hat{\Delta}^{+}_{{\bf k}}
\hat{\Delta}_{{\bf k}} ] = |{\bf d}({\bf k}) |^2 $. Obviously the two 
states are nodeless on the corresponding Fermi surfaces. Note that the
example for three dimensions corresponds to the Balian-Werthammer
state or the B-phase of superfluid $^3$He.\cite{LEGGETT}

We now consider the case of as quasi-two-dimensional system which is
characterized by the fact that the Fermi surface is open in one
direction, the $ z$-axis. 
In such a system the weak coupling approach does not
lead to a unique state, but we can find six degenerate states
with the same nodeless gap. In Sr$_2$RuO$_4$ their degeneracy is
lifted by spin-orbit coupling and the corresponding states labeled
according to the representation of the tetragonal crystal point group 
of this compound is given in Table 1.\cite{RICE-SIG,NG-SIG} We can
distinguish two types of 
states here: those which have $ {\bf d} $-vector that changes
orientation for different points on the Fermi surface belonging to the 
one-dimensional representation $ A_{1u}, A_{2u}, B_{1u} $ and $ B_{2u} 
$ and those which have a fixed $ {\bf d} $-vector orientation but
a finite orbital angular momentum, belonging to the two-dimensional $
E_u $ representation.\cite{RICE-SIG, NG-SIG} Note that the latter 
is the chiral state, i.e. it breaks time reversal symmetry and parity.
Since all these states are degenerate in the spin rotation symmetric case
beyond simple spin rotation transformation, the question arises which
among them is stable. For Sr$_2$RuO$_4$ experiments
suggest the chiral state with $ {\bf d}
\parallel \hat{{\bf z}} $.\cite{ISHIDA,LUKE} 

\begin{table}
\caption{Six-fold degenerated states in $p$-wave pairing symmetry}
\begin{indented} 
\item[]\begin{tabular}{@{}cc} 
\br 
$\Gamma$ &  ${\bf d}({\bf k})$ \\ 
\mr 
A$_{1u}$ &  $\hat{\bf x} \hat{k}_{x} + \hat{\bf y} \hat{k}_{y}$ \\ 
\mr
A$_{2u}$ &  $\hat{\bf x} \hat{k}_{y} - \hat{\bf y} \hat{k}_{x}$ \\ 
\mr
B$_{1u}$ &  $\hat{\bf x} \hat{k}_{x} - \hat{\bf y} \hat{k}_{y}$ \\ 
\mr 
B$_{1u}$ &  $\hat{\bf x} \hat{k}_{y} + \hat{\bf y} \hat{k}_{x}$ \\ 
\mr 
E$_{u}$ (chiral states) 
&  $\hat{\bf z} (\hat{k}_{x} \pm i \hat{k}_{y})$ \\ 
\br 
\end{tabular}
\end{indented} 
\end{table} 

In Sr$_2$RuO$_4$ the loss of spin rotation symmetry by spin-orbit
coupling carries the main responsibility in picking the stable state. 
In this letter, however, we assume that the spin rotation symmetry is 
preserved in the normal state so that all states listed in 
Table 1 have the same transition temperature $ T_c $ as a solution of
the linearized weak coupling gap equation. In this case the degeneracy
must be lifted in a higher order process. A well-known concept
introduced by Anderson and Brinkman is the spin fluctuation feedback
mechanism.\cite{ANDERSON,VOLLHARDT} If paramagnon exchange plays a
dominant role in the pairing  
interaction, the modification of the spin fluctuation spectrum by the
superconducting condensation also alters the pairing interaction. 
It was shown that this mechanism works in favor of the
so-called AMB-state or A-phase in $^3$He.\cite{LEGGETT} This mechanism
applied to 
the 2D situation turns out to stabilize the time reversal breaking
state which is indeed the analogue to the A-phase.\cite{SEOUL,HERA}

Here we would like to introduce an additional feedback mechanism which 
does not exist in a neutral Fermi liquid such as $^3$He. It is based
on the presence of chirality in the orbital part of pairing state 
and we will call it, therefore, {\it chiral
feedback mechanism}. It was shown that in the state $ {\bf d}({\bf k})=
{\bf n} (k_x \pm i k_y) $ a Chern-Simons-like term, 
\begin{equation}
\epsilon_{ij} (A_{0} \partial_{i} A_{j} + A_{i} \partial_{j} A_{0}) 
\end{equation} 
appears in the effective low-energy field theory of a static 
quasi-two-dimensional system ($ i,j=x,y$ ).
\cite{VOLOVIK,GORYO-ISHI}
Consequently,  
charge fluctuations generate local 
magnetic field distributions ($z$-axis oriented) and current
fluctuations lead to transverse electric
field distributions, whose orientation depends on the
chirality. This property yields an additional (anomalous) pairing interaction
in the superconducting state which has selective power for chirality.
This can be seen by the
following simple picture. The magnetic field induced
by the charge of a quasiparticle acts via the Lorentz force on a
passing-by quasiparticle.\cite{MORINARI} This force is either attractive or
repulsive depending on  
which side the quasiparticle trajectory is located (Fig. 1). In this way the 
interaction is attractive for one choice of Cooper pair angular
momentum (chirality) and repulsive for the other. The attractive interaction
appears for the same chirality realized in the pairs of the
condensate. Hence this leads to a positive feedback for the chiral
state. 
This effect does not exist for the other states. 

\begin{figure}
\centerline{
\epsfysize=5cm\epsffile{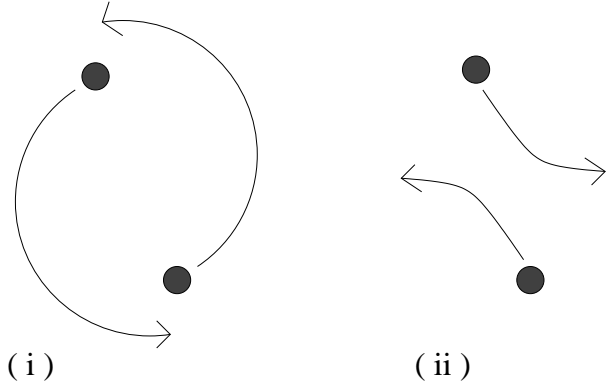}}
\caption{Two quasiparticles feel (i) attractive or (ii) repulsive 
interaction depending on their relative angular momentum. } 
\end{figure}

We will now discuss the effect by explicitly calculating its
contribution to condensation energy immediately below
the transition temperature $ T_c $. We represent the $ 
{\bf d} $-vector by $ {\bf d}({\bf p})= {\bf n}_{\gamma} d_{\gamma
i} k_{i}/k_{\rm F} $ where $ d_{\gamma i} $ is a complex order
parameter and $ k_{\rm F} $ inverse of the Fermi wavelength.
The band structure is simply parabolic $ \epsilon_{{\bf k}} = \hbar^2 (k_x^2 + 
k_y^2 - k_{\rm F}^2) / 2m_e $ without any dispersion along the $ z $-axis
leading to a cylindrical (open) Fermi surface. We assume that the
system has layered structure as Sr$_2$RuO$_4$ with an interlayer
spacing $d$ leading to the density of states, $N(0) = m_e/2 \pi \hbar^2 
d $, at the Fermi level. 
The anomalous pairing interaction appearing in the superconducting phase is
connected with the density-current correlation function which for the
2D electrons has the form (in the unit $\hbar=c=1$)
\begin{eqnarray}
\pi_{0j}(i \nu_{n}, {\bf q})&=& \int d \tau d^3 x e^{i
\nu_n \tau} e^{ i {\bf q} \cdot {\bf x}} \langle \hat{\rho}({\bf x}, \tau) 
\hat{j}_j (0,0) \rangle 
\nonumber \\
& = & k_B T \sum_{m} 
\int\frac{d^{3}k}{(2 \pi)^{3}} 
\frac{-(2 k_j + q_j)}{2m_e} 
{\rm Tr}[{\cal{G}}(i \omega_{m} + i
\nu_{n}, {\bf k} + {\bf q}) 
{\cal{G}}(i \omega_{m},{\bf k})
-
\nonumber \\
&&{\cal{F}}^{\dagger}(i \omega_{m} + i
\nu_{n}, {\bf k} + {\bf q}) 
{\cal{F}}(i \omega_{m},{\bf k})
]  ~~(j=1,2),   
\label{den-cur}
\end{eqnarray}
and the Green's functions are
\begin{equation}
{\cal {G}}(i \omega_{m},{\bf k})=\frac{i \omega_{m} + \epsilon({\bf
k})}{\omega_{m}^{2} + E ({\bf k})^{2}} \quad \mbox{and} \quad
{\cal {F}}(i \omega_{m},{\bf k})=-\frac{i \Delta({\bf
k})}{\omega_{m}^{2} + E({\bf k})^{2}}
\end{equation}
with $ E({\bf p})=\pm\sqrt{\epsilon({\bf k})^{2} + \Delta ({\bf
k})^2}$ and $\omega_{m}=(2 m + 1) \pi k_B T$ and $\nu_{n}=2 n \pi k_B T$ 
as the fermionic and bosonic Matsubara frequencies, respectively. 
We express $ \pi_{0j}(i \nu_{n}, {\bf q}) = i \epsilon_{ij} q_{i}
f(i \nu_{n},{\bf q}) + \nu_n q_j \pi(i \nu_n, {\bf q})$ due to
the gauge invariance. $f(i \nu_n, {\bf q})$ comes from the chirality 
and is written as 
\begin{equation}
f(i \nu_n, {\bf q}) = - \frac{i}{2!} \epsilon_{ij} \frac{\partial
\pi_{0j}(i \nu_n, {\bf q})}{\partial q_i}. 
\end{equation}
Close to $ T_c $ we can restrict ourselves to
the leading contributions in $ d_{\gamma i} $ and obtain, 
\begin{equation}
f(i \nu_{n},{\bf q}) = \frac{e^{2} k_B T_c}{2 m_e k^2_{F}}  
\sum_{m} \int \frac{d^{3} k}{(2 \pi)^{3}} 
\frac{- i \epsilon_{ij} d_{\gamma l} d_{\gamma l'}^{*}  
k_{l} k_{l'}} 
{\{(\omega_{m} + \nu_{n})^{2} + \epsilon({\bf k}+ {\bf q})^{2}\}
\{ \omega_{m}^{2} + \epsilon({\bf k})^{2} \}} 
\label{f-dyna}
\end{equation}
where we sum over repeated indices. Note that this expression is only
finite for a chiral p-wave state. 
We restrict ourselves to static
contributions and comment on the dynamical part later. 

We now consider the $q$-dependence. We
approximate $ \epsilon({\bf k} + {\bf q}) = \epsilon({\bf k}) + {\bf
v}_{\rm F} \cdot {\bf q} $ where $ {\bf v}_{F} $ is
the Fermi velocity. In order to evaluate the integral we introduce
cylindrical coordinates and perform first the integration over the
radial part and $ z$-component of $k$,
\begin{equation}
f({\bf q}) = \frac{i e^2 k_B T_c}{4 m_e} N(0) \pi (\epsilon_{ij}
d_{\gamma i} d_{\gamma l}^* ) \sum_m \int \frac{d
\theta}{2 \pi} \frac{\hat{k}_l \hat{k}_j }{|\omega_m|(|\omega_m|^2 +
\frac{1}{4} ({\bf v}_{\rm F}\cdot {\bf q})^2)}, 
\end{equation}
where $\hat{k}_i= k_i / |{\bf k}|$.   
For small $ q $ we may expand $ f ({\bf q}) $ as
\begin{equation}
f({\bf q}) \approx \frac{i e^2 N(0) \pi}{4 m_e (\pi k_B T_c)^2}
\epsilon_{ij} 
d_{\gamma i} d^*_{\gamma j} \left[ \frac{7}{4} \zeta(3) - 
\frac{31}{32} \zeta(5) \xi^2 q_{\perp}^2 + \cdots \right]
\end{equation}
where $ q_{\perp}^2 = q_x^2 + q_y^2 $, $ \xi = v_{\rm F}/ 2 \pi
k_B T_c $ defines the coherence length and $ \zeta(n) $ is the
zeta-function. The behavior of $ f({\bf q}) $ for $ \xi q_{\perp} \gg
1 $ is dominated by the regions of the $ \theta $-integral for which $ 
| {\bf v}_{\rm F} \cdot {\bf q} | \ll 2 \pi k_B T_c $ and leads to
\begin{equation}
f({\bf q}) \approx \frac{i e^2}{4 m_e} \frac{N(0)}{(\pi k_B T_c)^2}
\frac{\epsilon_{ij}
d_{\gamma i} d^*_{\gamma j}}{\xi q_{\perp}}. 
\end{equation}
Matching the limiting behaviors together we can approximate $ f({\bf
q}) $ by the following form,
\begin{equation}
f({\bf q}) \approx 
\frac{i e^2}{4 m_e} \frac{N(0) \pi}{(\pi k_B T_c)^2} 
\frac{\epsilon_{ij}
d_{\gamma i} d^*_{\gamma j}}{\sqrt{1 + \gamma \xi^2 q_{\perp}^2}}, ~
\gamma={\cal {O}}(1),  
\end{equation}
which represents the form factor of parity and time reversal symmetry 
breaking part in $ \pi_{0j}(0, {\bf q}) $. 

The current-charge density interaction introduced via $ \pi_{0j} (0,
{\bf q}) $ gives an additional contribution to the pairing interaction 
below the superconducting transition. As a feedback effect this
appears in the GL free energy in a fourth-order correction expressed
by
\begin{equation} 
\Delta F_{{\rm fb}} = k_B T_c \int \frac{d^{3} q}{(2 \pi)^3} D_{00}({\bf q}) 
\pi_{0i} ({\bf q}) D_{ij}({\bf q}) \pi_{j0}({\bf q}),   
\label{fd-c}
\end{equation} 
following the diagram in Fig.2. Here $D_{00}$, $D_{ij} ~(i,j=1,2)$ is
the gauge field propagators which in Coulomb gauge are,
\begin{equation} 
D_{00}({\bf q})=\frac{1}{q^{2} + l_{\rm TF}^{-2}} \quad \mbox{and}  
\quad D_{ij}({\bf q})= \frac{-\delta_{ij}}{q^{2}},  
\end{equation}  
These propagators contain all renormalizations, i.e. Thomas-Fermi
screening for the scalar potential with the screening length $ l_{\rm
TF} $. Since $ T \approx T_c $ London screening of the superconductor
can be neglected. We ignore here also the dynamical part for
simplicity, as it would give the same contributions for all competing
states. 

\begin{figure}
\centerline{
\epsfysize=5cm\epsffile{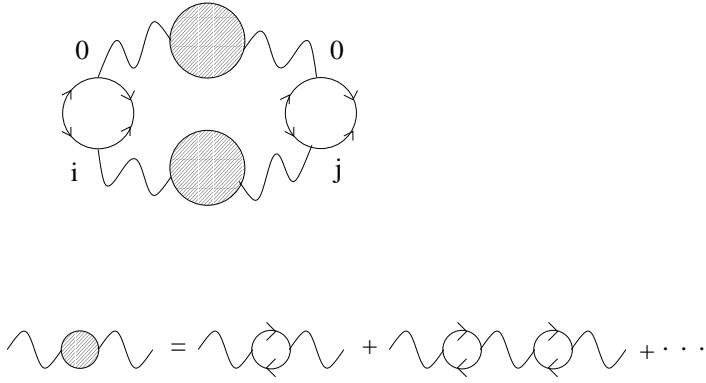}}
\caption{The diagram for the fourth order correction in GL 
free energy. The shadowed circles show the renormalization by the 
normal fermionic Green function.} 
\end{figure}

If we separate the $ q$-integration in $q_z $- and $ q_{\perp} $-part, 
we obtain,
\begin{eqnarray} 
\Delta F_{\rm fb}& = & \left\{ \frac{ e^2 N(0) \pi}{4 m_e (\pi k_B T_c)^2} 
\epsilon_{ij} d_{\gamma i} d_{\gamma j}^* \right\}^2 k_B T_c l_{\rm
TF}^2 
\nonumber \\
&& \quad \times \int \frac{d^2 q_{\perp}}{(2 \pi)^2} \int \frac{d q_z}{2 \pi}
\frac{q_{\perp}^2}{1 + \gamma \xi^2 q_{\perp}^2} \frac{1}{q_{\perp}^2 +
q_z^2} \frac{1}{1 + l_{\rm TF}^2 (q_{\perp}^2 + q_z^2)} 
\end{eqnarray}
After performing the $ q_z $-integration the remaining $ q_{\perp}
$-integral has a cutoff $ q_{\perp} \sim l_{\rm TF}^{-1} $. This leads 
to the free energy correction,
\begin{equation}
\Delta F_{\rm fb} 
\approx \frac{8 \alpha^2}{\pi} \frac{T_c}{T_{\rm F}} \frac{l_{\rm TF}}{d} 
\frac{N(0)}{(\pi k_B T_c)^2} (\epsilon_{ij} d_{\gamma i} d_{\gamma
j}^*)^2
\end{equation}
where we give an expression formally close to the conventional
fourth-order terms in order to give a comparison of its magnitude. 
Here we recover the constants $\hbar$ and $c$, the
factor $ \alpha = e^2 / \hbar c $ is the fine structure constant and the 
ratio $ T_c / T_{\rm F} $ indicates the strong coupling nature of the
correction term, similar to the spin fluctuation feedback
mechanism. \cite{VOLLHARDT} 

For the chiral $ p $-wave state $ \epsilon_{ij} d_{\gamma i} d_{\gamma
j}^* = i 2 \chi |\Delta | $ and zero for all other states ($ \chi =
\pm 1 $ denotes the chirality). Thus, the correction to the fourth order term
is negative definite and favors the chiral $p$-wave state. 
The ratio between this correction and the usual fourth order coefficient is
\begin{equation}
\frac{\delta \beta_{\rm fb}}{\beta} \sim \frac{\alpha^2}{\pi} 
\frac{T_c}{T_{\rm F}} \frac{l_{\rm TF}}{d} 
\end{equation}
which for Sr$_2$RuO$_4$ is of the order $ 10^{-6} $. 

It is easy to see that the dynamical contributions, taking into
account $ \nu_n \neq 0 $, does not change the result qualitatively. 
The corresponding coefficient in the free energy, however, increases. We 
have verified numerically that an increase of one order of magnitude
is possible. It is clear that other mechanisms, such as the spin fluctuation 
feedback or spin-orbit coupling, would dominate over the
chiral feedback effect in stabilizing the chiral $ p $-wave state. 
We would like to emphasize, however, that our analysis shows that
for a quasi-two-dimensional $ p $-wave superconductor the chiral
feedback effect, based on the anomalous coupling between charge and
current, supports the chiral superconducting phase and, thus, works in 
the same direction that the spin fluctuation feedback mechanism.
 
The authors are grateful to A. Furusaki, K. Ishikawa and T. Morinari 
for many helpful discussions. This work has been financially supported 
by a Grant-in-Aid of the Japanese Ministry of Education, Science,
Culture and Sports.


\end{document}